\documentclass[tightenlines,aps,
twocolumn,
nofootinbib
]{revtex4}
\newif\ifpdf\ifx\pdfoutput\undefined\pdffalse\else\pdfoutput=1\pdftrue\fi
\newcommand{\pdfgraphics}{\ifpdf\DeclareGraphicsExtensions{.pdf,.jpg}\else\fi}

\usepackage{float,axodraw}
\usepackage{psfig,float}
\usepackage{graphics}
\usepackage{amsmath}
\newcommand{\be}{\begin{equation}}
\newcommand{\ee}{\end{equation}}
\newcommand{\ba}{\begin{eqnarray}}
\newcommand{\ea}{\end{eqnarray}}

\begin{document}
\pdfgraphics
\begin{titlepage}

\begin{flushright}
\vbox{
\begin{tabular}{l}
UH-511-1086-06\\
 hep-ph/0604205
\end{tabular}
}
\end{flushright}

\vspace{0.6cm}

\title{ On the QCD corrections to Vainshtein's  theorem
for ${\bf \langle VVA \rangle} $ correlator
}

\author{Kirill Melnikov \thanks{
e-mail:  kirill@phys.hawaii.edu}}
\affiliation{Department of Physics and Astronomy,\\ University of Hawaii,\\
Honolulu, HI, D96822}


\begin{abstract}

\vspace{2mm}

We point out that, contrary to existing claims of the opposite, 
Vainshtein's theorem on the non-renormalization 
of the correlator  of an axial and two vector currents, 
is only valid in the chiral limit. When quarks, 
that contribute to the correlator,  are massive, 
the QCD corrections 
do not vanish and the transversal and longitudinal 
functions are renormalized differently. We compute those
corrections and study their implications for QCD effects 
in electroweak corrections to the muon anomalous magnetic moment.

\end{abstract}

\maketitle

\thispagestyle{empty}
\end{titlepage}

\section{Introduction}

A correlator of an axial and two vector currents $\langle VVA \rangle $
is a peculiar object. Because 
of  its relation to the anomaly of the axial current, 
it plays  important role in   understanding subtle issues in 
quantum field theory and  in certain  phenomenological applications. 

\begin{figure}[htb]
\centerline{
\psfig{figure=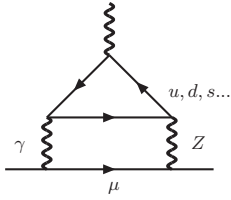,width=1.2in}
}
   \caption{
\label{fig1}
A two-loop diagram that contributes to electroweak corrections
 to the muon anomalous 
magnetic moment and  involves the anomalous $\langle VVA \rangle$ 
correlator.}
\end{figure}

One such application was recently identified in the physics of the 
muon anomalous magnetic moment where calculation 
of  two-loop electroweak corrections requires the 
correlator $\langle VVA \rangle $, Fig.\ref{fig1}. Thanks to  
Furry theorem, the vector part of the $Z$ coupling to fermions does 
not contribute and the diagram Fig.\ref{fig1} is entirely determined 
by the $\langle VVA \rangle $ correlator. 
Because of  
significant  interest in the physics of the muon anomalous magnetic 
moment, driven by   recent measurements \cite{g-2recent} 
of this observable at Brookhaven National Laboratory, it 
is not surprising that the two-loop electroweak corrections  to the muon 
magnetic anomaly were studied thoroughly. In particular, 
the influence of hadronic effects on electroweak corrections 
to the muon anomalous magnetic moment 
was discussed  in a number of recent papers \cite{CMV,knecht}. 
While these activities 
resulted in a better understanding of hadronic effects in the 
physics of the muon anomalous magnetic moment, 
they  also lead to a discovery of an interesting theoretical 
result -- a 
novel non-renormalization theorem for the $\langle VVA \rangle$
correlator \cite{nonren} (see also \cite{knecht1}). 

The content of this theorem can be explained as follows. 
In a special kinematic configuration when one of the vector currents 
is soft, the correlator of an axial and two vector currents
is described by the longitudinal and 
transversal functions $w_{\rm L,T}$. 
In the chiral limit, $w_{\rm L}$ does not receive 
QCD corrections as a consequence of the Adler-Bardeen theorem on 
the non-renormalization of the axial anomaly \cite{adler}. 
The new result, discovered 
by Vainshtein \cite{nonren}, is  that in the chiral limit 
the transversal function  is also not renormalized 
by perturbative QCD effects.  The proof is based on the observation 
that in the chiral limit $w_{\rm L} = 2w_{\rm T}$, 
to all orders in the strong coupling constant 
$\alpha_s$. Since 
the Adler-Bardeen theorem protects $w_{\rm L}$  from QCD corrections, 
the transversal function $w_{\rm T}$  is also not renormalized.

We note that these non-renormalization theorems \cite{nonren,knecht1}  
are formulated in  the chiral limit. 
Recently, Pasechnik and Teryaev  claimed \cite{ter} 
that these 
results  are more 
general and that even for massive quarks ${\cal O}(\alpha_s)$ 
QCD corrections to the longitudinal and transversal 
 functions are absent. This result 
looks somewhat peculiar;  for example, it is difficult to reconcile
the absence of ${\cal O}(\alpha_s)$ corrections  
with other well-known properties \cite{CMV,nonren} of the  
correlator $\langle VVA \rangle$ , 
in particular its operator product expansion. 
Intrigued  by that, we decided 
to re-calculate the QCD corrections to the $\langle VVA \rangle $ correlator 
in the limit when one of the vector currents is soft, allowing for 
non-zero quark masses. 
In contrast to \cite{ter}, we obtain non-vanishing QCD corrections 
away from the chiral limit. Below the details of the calculation 
are described.

\section{Calculation}
\label{calc}

Consider the correlator of an axial and two vector  currents 
\begin{equation}
T_{\mu \gamma \nu}^q = 
- \int {\rm d}^4 x {\rm d}^4 y e^{iqx - iky} 
\langle 0 | T \{ j_\mu(x) j_\gamma(y) j_\nu^5(0) \} | 0 \rangle,
\end{equation}
where 
\begin{equation}
j_\mu = Q_q \bar q \gamma_\mu q,
~~~~j_\mu^5 = \bar q \gamma_\mu \gamma_5 q
\end{equation}
are the vector and axial currents and $Q_q$ is the 
electric charge of the quark $q$. The pole  mass of the quark is denoted 
by $m_q$ in what follows.

We assume that momentum $k$ is much smaller than $q,m_q$ and view the 
current $j_\gamma(y)$ as a source of soft photons. Then, we can 
write 
\begin{equation}
\begin{split} 
& T_{\mu \nu}^q = T^q_{\mu \gamma \nu} e^\gamma (k), \\
& T_{\mu \nu}^q
= i \int {\rm d}^4 x e^{iqx} 
\langle 0| T \{ j_\mu(x) j_\nu^5(0) \} | \gamma(k) \rangle,
\end{split}
\label{eq2}
\end{equation}
where $e^\gamma(k)$ is the polarization 
vector of a photon with momentum  $k$.

We are interested in the expansion of the tensor $T_{\mu \nu}^q$ 
in $k$. Such an expansion was constructed in \cite{CMV,nonren}. There 
it was shown that, through first order in $k$, $T_{\mu \nu}^q$ can 
be written 
\begin{equation}
\begin{split}
& T_{\mu \nu}^q = \frac{-i}{4\pi^2}
\left [ 
w^q_{\rm L}(q^2) q_\nu q^\sigma \tilde f_{\sigma \mu}
+w^q_{\rm T}(q^2) \left ( 
- q^2 \tilde f_{\mu \nu} 
\right.  \right.
\\[2mm]
\label{eq3}
&  \left. \left.
~~~~~~~~+ q_\mu q^\sigma \tilde f_{\sigma \nu} - 
q_\nu q^\sigma \tilde f_{\sigma \mu} \right ) 
\right ],
\end{split}
\end{equation}
where $\tilde f_{\mu \nu} = \epsilon_{\mu \nu \alpha \beta} k^\alpha e^\beta$.
The functions $w_{\rm L,T}^q$ parametrize 
Lorentz  structures contributing to $T_{\mu \nu}^q$,
that are longitudinal (transversal) with respect to momentum $q$.

\begin{figure}[htb]
\centerline{
\psfig{figure=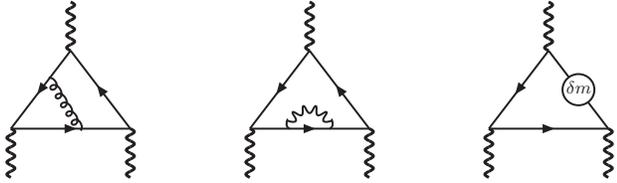,width=3.2in}
}
   \caption{Examples of two-loop diagrams that lead to ${\cal O}(\alpha_s)$ 
correction to $w^q_{\rm L,T}$.
The rightmost diagram represents an insertion of the mass counter-term.
\label{fig2}
}
\end{figure} 

In general, $w^q_{\rm L,T}$ depend on  $q^2$ and the quark mass $m_q$.
In the one-loop approximation, these 
functions were computed in \cite{rosenberg,kuhto}
\begin{equation}
\begin{split}
w^{q,0}_{\rm L} = 2 w^{q,0}_{\rm T} 
&
= 2 N_c Q_q^2 \int_{0}^{1}\!{\rm d}\xi \,
\frac{\xi(1-\xi)}{\xi(1-\xi)Q^{2}+m_q^{2}}\\[2mm]
&
=
\frac{2 N_c Q_q^2}{Q^{2}}\,
\Big(1-\frac{2m_q^{2}}{\beta Q^{2}}\ln\frac{\beta+1}{\beta -1}\Big)
\,.
\label{eq5}
\end{split}
\end{equation}
In Eq.(\ref{eq5}),  $N_c = 3$ is the number of colors. We also introduced 
the  Euclidean momentum $Q$, $Q^2 = -q^2$ and 
$\beta = \sqrt{1+4m_q^2/Q^2}$.  
In the limit $m_q=0$, we find
$w^{q,0}_{\rm L} = 2w^{q,0}_{\rm T} = 2N_c Q_q^2/Q^2$; as follows 
from the Adler-Bardeen and Vainshtein's 
theorems \cite{adler,nonren}, this is exact perturbative
result in that all QCD corrections to it vanish.

In this paper, we study ${\cal O}(\alpha_s)$ QCD corrections 
to $w^{q,0}_{\rm L,T}$ in case when the quark mass is not zero.
We have to compute twelve  two-loop diagrams; some examples are 
shown in Fig.\ref{fig2}. Because computing $w_{\rm L,T}^q$ 
beyond the one-loop approximation for arbitrary value of $m_q^2/q^2$ 
is technically challenging, we  
adopt a different approach. It turns out that the calculation is 
drastically simplified in two limiting cases $Q \ll m_q$ and $Q \gg m_q$. 
In addition, if sufficiently many terms in the 
expansion of $w_{\rm L,T}^q$ in these limits are known, we can 
reconstruct the two functions completely using Pad\'e approximation. 

For both cases, $Q \ll m_q$ and $Q \gg m_q$, we employ the 
method of asymptotic expansions \cite{smirnov}.
In the large-mass 
limit, $m_q \gg Q$, 
the computation is particularly simple since Taylor expansion of Feynman 
diagrams in Fig.\ref{fig2} in both $k$ and $Q$ suffices. Consequently, 
only two-loop vacuum Feynman integrals have to be evaluated in this case.

The small mass limit $m_q \ll Q$ is more complex, since   Taylor 
expansion in $m_q$ is insufficient. 
This can 
be seen from the one loop result which is non-analytic in the limit 
$m_q \to 0$. To expand the two-loop diagrams, Fig.\ref{fig2}, in 
$m_q$, we have to consider how 
a loop momentum $l$, flowing along a line in a diagram, compares 
with $m_q$ and $Q$. If $l \sim Q$, the propagator is Taylor expanded in $m_q$.
In the opposite case, $l \sim m_q$, the propagator $1/((q+l)^2-m_q^2)$ is expanded both in $l$ and $m_q$ while the propagator $1/(l^2-m_q^2)$ is left unexpanded.  

For each diagram, we have to consider different 
 momenta routings and 
identify all possible subgraphs. Among them, there are two types of 
subgraphs that can be easily described. The first one corresponds to the 
situation when all lines in a diagram are off-shell by an amount of 
order $Q$. In this case, Taylor expanding diagram in $m_q$ leads to 
a massless computation. The opposite situation occurs when 
both loop momenta are of order $m_q$. Then, a diagram is expanded 
in $1/Q$ and the integrals we have to deal with are the two-loop massive 
vacuum integrals. In  intermediate cases when some of the lines 
in a diagram are hard $\sim Q$  
and other are soft $\sim m_q$, the two-loop graph factorizes into the product 
of one-loop graphs.

To compute the transversal and longitudinal functions separately, 
we choose different 
momenta of the soft photon. 
For example, if we choose the soft momentum $k$ 
in such a way that $k \propto  q$,
we project out the longitudinal 
structure function. To get rid of  free Lorentz indices in that case, 
we contract the tensor $T_{\mu \gamma \nu}$, defined in Eq.(\ref{eq2}), 
with $e^{\mu \gamma \nu  \beta} q_\beta/\sqrt{-q^2}$.  
This procedure allows us to deal with  scalar Feynman two-loop integrals
for the computation of $w^q_{\rm T}$. As we pointed out in the previous 
paragraph,  the calculation of these integrals is simple once 
the limits $m_q \ll Q$ or $m_q \gg Q $ are considered. 
A similar procedure is used to compute $w^q_{\rm L}$.

We use dimensional regularization for the computation. For the 
 consistent 
treatment  of the axial current in $d$ dimensions, 
we employ the procedure of  Ref.\cite{larin}.  The calculation is 
performed in an arbitrary covariant gauge; the cancellation of the 
gauge parameter dependence  in the final result is a welcome check on the 
correctness of the calculation. To present the results for the longitudinal
and transversal functions, we define 
\begin{equation}
w^q_{\rm L,T} = \frac{N_c Q_q^2}{Q^2}
\left ( \Delta_{\rm L,T}^{q,(0)} 
+ \frac{C_F \alpha_s}{\pi} \Delta_{\rm L,T}^{q,(1)}
+{\cal O}(\alpha_s^2) \right ).
\end{equation}
For  $Q \ll m_q$, we find
\begin{equation}
\begin{split}
& 
\Delta^{q,(1)}_{\rm L} = \frac{Q^2}{3m_q^2}
\left (
\frac{7}{2} - \frac{617}{540}\frac{Q^2}{m_q^2} 
+\frac{8041}{25200} \frac{Q^4}{m_q^4}\right )
+...,
\\
& 
\Delta^{q,(1)}_{\rm T} = 
\frac{Q^2}{6m_q^2} 
\left (
3 - \frac{521}{540}\frac{Q^2}{m_q^2} + 
\frac{449}{1680} \frac{Q^4}{m_q^4} \right ) +..,
\end{split}
\label{eq15}
\end{equation}
where 
ellipses stands for higher order terms in the expansion in 
$Q^2/m_q^2$.  In case $Q \gg m_q$, the functions read 
\begin{equation}
\begin{split}
& 
\Delta^{q,(1)}_{\rm L} = 
\frac{m_q^2}{Q^2} 
\left ( 5 L^2 +  L + \frac{19}{6} - 8 \zeta(3)
\right )
\\
& +\frac{m_q^4}{Q^4} \left ( -\frac{58 L^2}{3}  + \frac{286 L}{9}
+ \frac{421}{27}  \right )
+...,
\\
& \\
& 
\Delta^{q,(1)}_{\rm T} = 
\frac{m_q^2}{Q^2} 
\left ( \frac{5 L^2}{2} - \frac{L}{2}  +\frac{13}{12} - 4 \zeta(3)
\right 
)
  \\
&  +\frac{m_q^4}{Q^4} \left (- \frac{19L^2}{3} + \frac{145L}{9}
+ \frac{335}{54}  \right ) +...,
\end{split}
\label{eq6}
\end{equation}
where $L = \ln(Q^2/m_q^2)$, $\zeta(3)$ is 
the Riemann zeta-function  and ellipses stands 
for other terms suppressed by higher power of $m_q^2/Q^2$.

Several things can be pointed out in connection with these results. 
As expected,  the QCD corrections to $w^q_{\rm L,T}$ 
vanish in the limit $m_q \to 0$, in accord with the non-renormalization 
theorems \cite{adler,nonren}. 
Also, the QCD corrections are not universal so that
the all-orders perturbative relation $w^q_{\rm L} = 2 w^q_{\rm T}$
is violated, once non-zero quark masses are allowed.

\begin{figure}[tb]
\centerline{
\psfig{figure=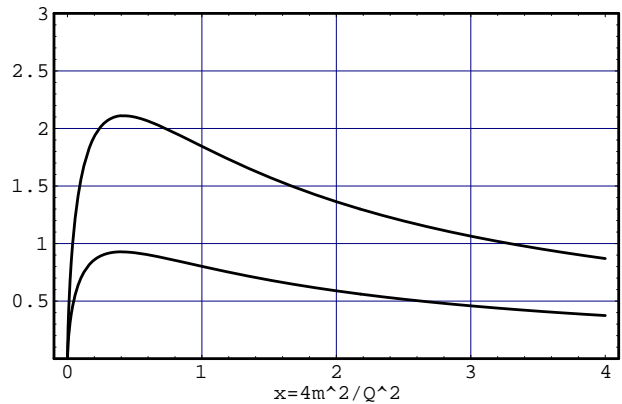,width=3.2in}
}
   \caption{The functions $\Delta_{\rm L}^{\rm q,(1)}$ (upper curve) and 
$\Delta_{\rm T}^{\rm q,(1)}$ (lower curve) plotted 
in dependence on $x = 4m_q^2/Q^2$.
\label{fig4}
}
\end{figure}

Using  Eqs.(\ref{eq15},\ref{eq6}), 
we can  construct  the entire functions $\Delta^{q,(1)}_{\rm L,T}$ 
with the help of  Pad\'e approximation \cite{pade}.  To this end, 
we compute ten terms in the expansion of $\Delta^{q,(1)}_{\rm L,T}$ 
in  $Q^2/m_q^2$ 
and perform Pad\'e approximation replacing the series by rational 
functions. It turns out that the resulting functions do not depend
on the precise form of the Pad\'e approximation employed for their 
construction.
For computations described below we use 
the $[4/5]$ Pad\'e approximant.

The virtue of the Pad\'e approximation is that it allows us to continue the 
small-$Q$ expansion of the functions $\Delta_{\rm L,T}^{q,(1)}$
to large  values of $Q$,  where it should smoothly merge  
with the large-$Q$ expansions of these functions Eq.(\ref{eq6}). 
Indeed, we observe a perfect match of the Pad\'e approximants 
and the large-$Q^2$ expansion of the functions $\Delta^{q,(1)}_{\rm L,T}$ 
in the interval  $ 10m_q^2 < Q^2 < 100 m_q^2$. Hence, the  Pad\'e 
approximants,  supplemented with large-$Q^2$ asymptotic, can be used to 
deduce $\Delta_{\rm L,T}^{q,(1)}$  
for arbitrary values of $Q^2/m_q^2$.

For completeness, we present below the functions $\Delta^{q,(1)}_{\rm L,T}$ 
in the 
form of the $[4/5]$ Pad\'e approximants. We introduce 
$z = Q^2/m^2$ and write
\begin{equation}
\Delta_{\rm L,T}^{q,(1)} = 
\frac{\sum \limits_{i=1}^{4} a_{\rm L,T}^{(i)}z^i}{1 +\sum \limits_{i=1}^{5} b_{\rm L,T}^{(i)} z^{i}}.
\end{equation}
The coefficients $a_{\rm L,T}^{(i)},b_{\rm L,T}^{(i)}$ are given in 
Tables~\ref{table1},\ref{table2}. The functions $\Delta^{q,(1)}_{\rm L,T}$ 
are plotted in  Fig.\ref{fig4}. It is interesting  that 
for large range of $m_q^2/Q^2$, the approximate relation 
$\Delta_{\rm L}^{q,(1)} \approx 2.32 \Delta_{\rm T}^{q,(1)}$ holds 
with reasonable accuracy.

\begin{tiny}
\begin{table}[htbp]
\vspace{0.1cm}
\begin{center}
\begin{tabular}{|c|c|c|}
\hline\hline
       &  L  &  T \\ \hline \hline
$a^{(1)}$ & $1.1666666666$ & $0.5000000000$ \\ \hline
$a^{(2)}$ & $0.3762074740$ & $0.1620591351$ \\ \hline
$a^{(3)}$ & $0.03273978142$& $0.01418000768$ \\ \hline
$a^{(4)}$ & $ 6.50545897 \cdot 10^{-4}$ & $2.836663049 \cdot 10^{-4}$ 
\\ \hline \hline
\end{tabular}
\caption{\label{table1} Coefficients for the $[4/5]$ Pad\'e approximant for 
the longitudinal and transversal functions.
}
\vspace{-0.1cm}
\end{center}
\end{table}
\end{tiny}

\begin{tiny}
\begin{table}[htbp]
\vspace{0.1cm}
\begin{center}
\begin{tabular}{|c|c|c|}
\hline\hline
       &  L  &  T \\ \hline \hline
$b^{(1)}$ & $0.6489185756$ & $0.6457232084$ \\ \hline
$b^{(2)}$ & $0.1487376001$ & $0.1469404864$ \\ \hline
$b^{(3)}$ & $0.01420626517$& $0.01388076895$ \\ \hline
$b^{(4)}$ & $5.028490259 \cdot 10^{-4}$ & $4.828123552 \cdot 10^{-4}$ 
\\ \hline 
$b^{(5)}$ & $3.78544 \cdot 10^{-6}$ & $ 3.527956921 \cdot 10^{-6}$
\\ \hline \hline
\end{tabular}
\caption{\label{table2} Coefficients for the $[4/5]$ Pad\'e approximant for 
the longitudinal and transversal functions.
}
\vspace{-0.1cm}
\end{center}
\end{table}
\end{tiny}

\section{Relation to the OPE}

As we pointed out in the Introduction, the non-renormalization of the 
functions $w_{\rm L,T}^q$ by perturbative QCD effects beyond the chiral 
limit, claimed 
in \cite{ter}, is not possible 
to understand given the operator product expansion (OPE)
of the tensor $T_{\mu \nu}^q$ derived in \cite{CMV,nonren}.   
In this Section, we elaborate on this statement.

Typically, the OPE is employed to estimate non-perturbative corrections 
to, otherwise, nearly perturbative observables. This is achieved by  
separating   physics at different distance or momentum scales. Soft 
contributions, associated with non-perturbative effects, are characterized 
by momentum scale $\Lambda_{\rm QCD}$ whereas hard contributions are 
characterized by  external kinematic scale $Q \gg \Lambda_{\rm QCD}$.
When the OPE is applied, soft components are identified with matrix elements 
of local operators while hard momenta contribute to Wilson coefficients 
of these operators. 

Although the OPE is associated with computations beyond perturbation theory, 
its main idea, separation of physics at different distance scales, makes 
it suitable for {\it perturbative} computations if largely different scales 
are present. Hence, it is well-suited for studying 
longitudinal and transversal functions
in the limit $Q \gg m_q$. In this case, soft scale is set 
by the mass of the quark $m_q$ rather than $\Lambda_{\rm QCD}$, but the OPE 
remains intact.

The OPE 
of the longitudinal and transversal functions reads \cite{CMV,nonren}
\begin{equation}
w_{\rm L,T}^q = \sum \limits_{i}^{} c_{\rm L,T}^i(\mu) \kappa_i(\mu),
\label{eq.ope}
\end{equation}
where the normalization  scale $\mu$ is introduced to make the 
separation of hard and soft modes unambiguous.
In Eq.(\ref{eq.ope}), the sum includes all local operators that contribute 
to the matrix element $\langle 0 | {\cal O}^i | \gamma(k)\rangle $ 
and behave as rank two pseudo-tensors under Lorentz transformations. 
The 
matrix elements are parametrized by 
\begin{equation}
\langle 0| {\cal O}_{\alpha \beta}^{i}(\mu) | \gamma(k) \rangle 
= -\frac{i}{4\pi} \tilde f_{\alpha \beta} \kappa_i(\mu),
\end{equation}
where the dependence on the normalization point is made explicit.

From our results in the previous Section, it follows that the 
relation $w_{\rm L} = 2w_{\rm T}$ is violated once ${\cal O}(\alpha_s
m^2/Q^2)$ contributions to these functions are computed.
We would like to analyze this result from the OPE perspective.
According to \cite{CMV,nonren}, there are  two operators that 
are required to compute quadratic mass correction to the massless 
limit of the transversal and longitudinal functions.
The first one is the dimension-two  operator,
${\cal O}^F_{\alpha \beta} = \tilde F_{\alpha \beta}/(4\pi)$, where  
$\tilde F_{\alpha \beta} 
= \epsilon_{\alpha \beta \mu \nu}\partial^\mu A^\nu$;
its matrix element between the vacuum and the soft photon  
is trivial and 
leads to $\kappa_F = 1$. The second operator  
is the  dimension-three operator 
${\cal O}_{\alpha \beta}^q = -i \bar q \sigma_{\alpha \beta} \gamma_5 q$. 
Its leading order Wilson coefficient reads \cite{CMV,nonren}
\begin{equation}
c_{\rm L}^{q}(\mu) = 2 c_{\rm T}^q(\mu) = \frac{4 Q_q  m_q(\mu)}{Q^4},
\label{eq6.5}
\end{equation}
where $m_q(\mu)$ is the running quark mass 
\begin{equation}
m_q(\mu) = m_q \left 
[ 1 + \frac{C_F \alpha_s}{\pi} \left (
-\frac{3}{4} \ln \frac{\mu^2}{m_q^2} -1 \right ) \right
].
\label{eq6.6}
\end{equation}

From the OPE of the functions $w_{\rm L,T}^q$, Eq.(\ref{eq.ope}),
 it follows that, in order 
to compute these functions through ${\cal O}(\alpha_s m_q^2/Q^2)$,  
we require the Wilson coefficient of the operator 
${\cal O}^F_{\alpha \beta}$, the matrix element of the 
operator ${\cal O}_{\alpha \beta}^q$ and the Wilson coefficients 
$c_{\rm L,T}^q$ through ${\cal O}(\alpha_s)$. To obtain those ingredients, 
two- and one-loop computations are needed.
It turns out that  all the matrix elements and Wilson coefficients
can be extracted from the results
reported in the previous Section. This happens because 
our calculation is based on computing contributions from different 
momenta regions separately. 
Then, soft regions in the one-  and  two-loop cases 
are identified with the matrix element of 
the operator ${\cal O}_{\alpha \beta}^q$ through 
${\cal O}(\alpha_s)$. We have checked that identical results 
for this matrix element are obtained, independent of whether 
such identification is done  
in $w_{\rm L}^q$ or $w_{\rm T}^q$. This independence is 
simple, yet welcome, 
check on the correctness of the calculation.

\begin{figure}[htb]
\centerline{
\psfig{figure=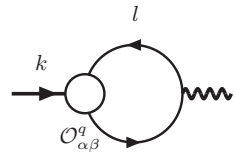,width=1.2in}
}
   \caption{The one-loop contribution to the matrix element 
of $\langle 0| {\cal O}_{\alpha \beta}^q | \gamma(k) \rangle$.
The two-loop contribution is obtained by adding all possible 
single gluon exchanges to this diagram.
\label{fig3}
}
\end{figure}

The matrix element $\langle 0| {\cal O}_{\alpha \beta}^q|\gamma \rangle$
is divergent (cf. Fig.\ref{fig3}). To arrive at the renormalized matrix 
element we use the $\overline {\rm MS}$ renormalization scheme and the 
fact that, under renormalization, ${\cal O}_{\alpha \beta}^q$ mixes 
with the operator $m \tilde F_{\alpha \beta}$. We obtain 
\begin{equation}
\begin{split}
& \kappa_q(\mu) = N_c Q_q m_q(\mu)
\left [ 
-L_\mu + \frac{8}{3} 
\right. \\
& \left. + 
\frac{C_F \alpha_s(\mu)}{\pi}
\left (-\frac{L_\mu^2+L_\mu}{4}
+ \frac{119}{48} + \frac{\pi^2}{8} 
\right )
\right ],
\label{eq.kappa}
\end{split}
\end{equation}
where $L_\mu = \ln(\mu^2/m_q^2(\mu))$.

Using Eq.(\ref{eq.kappa}), we determine Wilson coefficients of the 
operators ${\cal O}_{\alpha \beta}^{F,q}$ by matching the 
OPE of the functions $w_{\rm L,T}^q$, Eq.(\ref{eq.ope}), to perturbative 
result Eq.(\ref{eq6}). 
For simplicity, we give those Wilson coefficients at the 
normalization scale $\mu=Q$, 
where they do not contain any logarithms. To present the results, 
we introduce a short-hand notation $\bar m_q = m_q(Q)$ and 
\begin{equation}
c_{\rm L,T}^F = \frac{N_c Q_q^2}{Q^2}\; \tilde c_{\rm L,T}^F.
\end{equation}
Through ${\cal O}(\alpha_s)$, we find
\begin{equation}
\begin{split}
& \tilde c_{\rm L}^F = 2 - \frac{{\bar m_q}^2}{Q^2} \left [
\frac{32}{3} 
- \frac{C_F \alpha_s}{\pi} \left (\frac{5}{4} - \frac{\pi^2}{2} - 8 \zeta_3 
\right ) \right ], \\
& \tilde c_{\rm T}^F = 1 - \frac{{\bar m_q}^2}{Q^2} \left [
\frac{16}{3} 
- \frac{C_F \alpha_s}{\pi} \left (-\frac{61}{24} 
- \frac{\pi^2}{4} - 4 \zeta_3 
\right ) \right ],
\end{split}
\end{equation}
and 
\begin{equation}
\begin{split}
& c_{\rm L}^q = \frac{4 Q_q \bar m_q}{Q^4}, 
\\
& c_{\rm T}^q = \frac{2 Q_q \bar m_q }{Q^4} \left (1 + \frac{C_F \alpha_s}{2 \pi}
\right ).\\
\end{split}
\end{equation}

\section{The muon anomalous magnetic moment}

The correlator of an axial and two vector currents 
in the kinematic limit when one of the 
vector currents is soft, appears in the physics of the muon anomalous 
magnetic moment, Fig.\ref{fig1}. The correction to the muon magnetic anomaly 
due to these diagrams is obtained \cite{CMV} upon integrating 
$w_{\rm L,T}^q$ over $Q^2$ with the weight function
\begin{equation}
\Delta a_\mu = \frac{\alpha}{\pi} \frac{G_\mu m_\mu^2}{8\pi^2 \sqrt{2}}
\int \limits_{m_\mu^2}^{\infty} 
{\rm d}Q^2 \left ( w_{\rm L} + \frac{m_Z^2}{m_Z^2 + Q^2} w_{\rm T} 
\right).
\label{eq20}
\end{equation}
In Eq.(\ref{eq20}) we use
\begin{equation}
w_{\rm L,T} = \sum_{f} 2I_3^f w_{\rm L,T}^f,
\end{equation}
and the summation index $f$ denotes both lepton and quark contributions 
to the $\langle VV A \rangle $ correlator. In addition,  $I_3^f$ 
stands for  the weak isospin of the  fermion $f$.

In Section~\ref{calc} we derived  $w^q_{\rm L,T}$, including 
${\cal O}(\alpha_s)$ QCD corrections. We can use those results 
in Eq.(\ref{eq20}) to obtain 
QCD corrections to  $\Delta a_\mu$. 
Note that, at large $Q$,  the functions $\Delta_{\rm L,T}^{q,(1)}$ 
decrease as  $m_q^2/Q^2$; hence, at order ${\cal O}(\alpha_s)$ 
we can compute the integral in Eq.(\ref{eq20}) for each quark flavor 
separately. 
We only include charm, bottom and top 
quarks in the computation. The up, down and strange  
quarks have to be treated non-perturbatively \cite{CMV}.

We use $m_c = 1.5~{\rm GeV}$, $m_b = 4.8~{\rm GeV}$ 
and $m_t= 180~{\rm GeV}$ for  charm, bottom and top quark masses.  
We also employ the leading order running of the strong 
coupling constant with $\alpha_s(m_c) = 0.3$.
We obtain the following QCD corrections due to charm, bottom and top 
quarks
\begin{equation}
\Delta a_\mu^{q,\alpha_s} 
= \frac{\alpha}{\pi} \frac{G_\mu m_\mu^2}{8\pi^2 \sqrt{2}}
\left \{ 
\begin{array}{cc}
1.80,\,\, & q=c,\\
-0.33,\,\, & q=b, \\
0.56,\, & q=t.
\end{array}
\right.
\label{eq21}
\end{equation}
The  perturbative QCD correction is given by the sum of entries 
in Eq.(\ref{eq21}). We obtain
\begin{equation}
\Delta a_\mu^{\alpha_s} = 0.55 \times 10^{-11}.
\end{equation}
This correction is  well within 
the error bars, assigned to hadronic uncertainties  in electroweak 
contributions to the muon magnetic anomaly
  in Ref.\cite{CMV}; it is negligible for phenomenology 
given that   the current uncertainty 
on the muon anomalous magnetic moment
is $\sim 100 \times 10^{-11}$.

\section{Conclusion}

In this paper we computed the QCD corrections to the 
longitudinal and transversal functions  $w^q_{\rm L,T}$ allowing 
for non-zero quark masses. These functions describe 
the anomalous $\langle VVA \rangle $ correlator in the 
kinematic limit when one of the vector currents is soft.

According to the Adler-Bardeen and 
Vainshtein's theorems, $w^q_{\rm L,T}$ do not receive QCD corrections 
in the chiral $m_q=0$ limit. Our study is motivated by the claim in 
Ref.\cite{ter} that the absence of  QCD corrections $w_{\rm L,T}^q$
is also valid beyond the chiral limit when quark masses are allowed.

We performed explicit calculation of the QCD corrections 
to $w^q_{\rm L,T}(Q)$ for non-zero quark masses in two   kinematic limits
$m_q \ll Q$ and $Q \ll m_q$ and used 
Pad\'e approximation to derive the approximate form of these 
functions valid for arbitrary relation between $Q$ and $m_q$.
In variance with results of  Ref.\cite{ter},
we  found that  $w^q_{\rm L,T}$
receive perturbative corrections if $m_q \ne 0$.
We  argued   that these 
QCD corrections are consistent with the well-known  
operator product expansion of the correlator 
$\langle VVA \rangle$. 
 We studied the implications of our results for
the QCD corrections to electroweak contributions
to the muon anomalous magnetic 
moment that contain the correlator of an axial current and two vector 
currents, Fig.\ref{fig1},  and found  negligible effect.

{\bf Acknowledgments} I am grateful to A.~Vainshtein for useful comments.
This research is partially supported 
by the DOE under grant number 
DE-FG03-94ER-40833, the DOE Outstanding Junior Investigator Award 
and by the Alfred P. Sloan Foundation.

\end{document}